\documentclass[12pt]{article}

\usepackage{scicite}
\usepackage{times}
\usepackage{amsmath}
\usepackage{url}
\usepackage{longtable}
\usepackage{amssymb}
\usepackage{graphicx}
\usepackage{pdflscape}
\usepackage{multirow}
\usepackage{color}

\def\lsim{\hbox{\rlap{\raise 0.425ex\hbox{$<$}}\lower 0.65ex\hbox{$\sim$}}}
\def\gsim{\hbox{\rlap{\raise 0.425ex\hbox{$>$}}\lower 0.65ex\hbox{$\sim$}}}

\def\arcsec{\hbox{$^{\prime\prime}$}}

\newcommand{\nat}{Nature}

\def\hst{{\it HST}}
\def\swift{{\it Swift}}

\def\fermi{{\it Fermi}}

\def\ucsc{1}
\def\lbl{2}
\def\berk{3}
\def\dark{4}
\def\car{5}
\def\haw{6}
\def\lco{7}
\def\stsci{8}
\def\jhu{9}
\def\nat{10}

\def\arcsec{\hbox{$^{\prime\prime}$}}

\newdimen\sa  \newdimen\sb
\def\parcs{\sa=.07em \sb=.03em
     \ifmmode $\rlap{.}$^{\scriptscriptstyle\prime\kern -\sb\prime}$\kern -\sa$
     \else \rlap{.}$^{\scriptscriptstyle\prime\kern -\sb\prime}$\kern -\sa\fi}

\newenvironment{sciabstract}{%
\begin{quote} \bf}
{\end{quote}}

\newcounter{lastnote}

\title{Electromagnetic Evidence that SSS17a is the Result of a Binary Neutron Star Merger} 

\author
{C.~D.~Kilpatrick$^{\ucsc}$,
R.~J.~Foley$^{\ucsc}$,
D.~Kasen$^{\lbl,\berk}$,
A.~Murguia-Berthier$^{\ucsc}$,\\
E.~Ramirez-Ruiz$^{\ucsc, \dark}$,
D.~A.~Coulter$^{\ucsc}$,
M.~R.~Drout$^{\car}$,
A.~L.~Piro$^{\car}$,\\
B.~J.~Shappee$^{\car,\haw}$,
K.~Boutsia$^{\lco}$,
C.~Contreras$^{\lco}$,
F.~Di~Mille$^{\lco}$,
B.~F.~Madore$^{\car}$,\\
N.~Morrell$^{\lco}$,
Y.-C.~Pan$^{\ucsc}$,
J.~X.~Prochaska$^{\ucsc}$,
A.~Rest$^{\stsci, \jhu}$,
C.~Rojas-Bravo$^{\ucsc}$,\\
M.~R.~Siebert$^{\ucsc}$,
J.~D.~Simon$^{\car}$,
N.~Ulloa$^{\nat}$
\\}

\date{}

\begin{document} 

\baselineskip24pt

\maketitle 

\noindent
\normalsize{$^{\ucsc}$Department of Astronomy and Astrophysics, University of California, Santa Cruz, CA 95064, USA}\\
\normalsize{$^{\lbl}$Nuclear Science Division, Lawrence Berkeley National Laboratory, Berkeley, CA 94720, USA}\\
\normalsize{$^{\berk}$Departments of Physics and Astronomy, University of California, Berkeley, CA 94720, USA}\\
\normalsize{$^{\dark}$DARK, Niels Bohr Institute, University of Copenhagen, Blegdamsvej 17, 2100 Copenhagen, Denmark}\\
\normalsize{$^{\car}$The Observatories of the Carnegie Institution for Science, 813 Santa Barbara Street, Pasadena, CA 91101}\\
\normalsize{$^{\haw}$Institute for Astronomy, University of Hawaii, 2680 Woodlawn Drive, Honolulu, HI 96822, USA}\\
\normalsize{$^{\lco}$Las Campanas Observatory, Carnegie Observatories, Casilla 601, La Serena, Chile}\\
\normalsize{$^{\stsci}$Space Telescope Science Institute, 3700 San Martin Drive, Baltimore, MD 21218, USA}\\
\normalsize{$^{\jhu}$Department of Physics and Astronomy, The Johns Hopkins University, 3400 North Charles Street, Baltimore, MD 21218, USA}\\
\normalsize{$^{\nat}$Departamento de F\'{i}sica y Astronom\'{i}a, Facultad de Ciencias, Universidad de La Serena, Cisternas 1200, La Serena, Chile}\\
\\ 
\normalsize{$^\ast$To whom correspondence should be addressed; E-mail: cdkilpat@ucsc.edu.}

\begin{sciabstract}
11~hours after the detection of gravitational wave source GW170817 by the Laser Interferometer Gravitational-Wave Observatory and Virgo Interferometers, an associated optical transient SSS17a was discovered in the galaxy NGC~4993. While the gravitational wave data indicate GW170817 is consistent with the merger of two compact objects, the electromagnetic observations provide independent constraints of the nature of that system. Here we synthesize all optical and near-infrared photometry and spectroscopy of SSS17a collected by the One-Meter Two-Hemisphere collaboration. We find that SSS17a is unlike other known transients.  The source is best described by theoretical models of a kilonova consisting of radioactive elements produced by rapid neutron capture (the $r$-process). We find that SSS17a was the result of a binary neutron star merger, reinforcing the gravitational wave result.
\end{sciabstract}

\section*{Introduction}
The gravitational wave event GW170817 \cite{Abbott17} was detected by the Laser Interferometer Gravitational-Wave Observatory (LIGO) \cite{Aasi15} and Virgo Interferometer (Virgo) \cite{Acernese15}. An electromagnetic counterpart, SSS17a, was subsequently identified 11 hours after the gravitational wave event \cite{Coulter17}. Combining the gravitational wave and electromagnetic signatures could inform models of the physical mechanisms involved in compact object mergers \cite{Li98} and the rate at which compact object binaries produce heavy elements \cite{Ramirez-Ruiz15}.

Two seconds after GW170817 and 11 hours before SSS17a was identified, the {\it Fermi Gamma-ray Space Telescope} (\fermi) and International Gamma-Ray Astrophysics Laboratory (INTEGRAL) detected a $\gamma$-ray burst (GRB), GRB~170817A, coincident with GW170817 \cite{Goldstein17,Savchenko17}. It has long been hypothesized that short-duration GRBs (SGRBs) are the result of compact binary mergers involving neutron stars and/or black holes \cite{Paczynski86}. Rapidly fading GRB afterglows and even kilonova emission powered by the decay of heavy elements have been hypothesized as potential counterparts to compact binary mergers \cite{Li98}.

As the LIGO/Virgo collaboration analyze GW170817 and the information encoded in the waveform of the gravitational signal, as astrophysicists we seek to place stringent constraints on the nature and evolutionary behavior of the progenitor system by interpreting the electromagnetic properties of SSS17a and its galactic environment. We take an agnostic approach motivated by the question: if we had no constraints from gravitational waves and SSS17a were discovered as part of a typical optical transient survey, what could we conclude about the nature of the progenitor system? Such considerations may be unavoidable given that the LIGO and Virgo interferometers will undergo temporary off-line periods. In addition, deep, wide-field surveys, such as the Large Synoptic Survey Telescope, are expected to be more efficient at detecting SSS17a-like objects than targeted follow-up of LIGO/Virgo detections \cite{Siebert17}.

SSS17a was identified by the One-Meter Two-Hemisphere (1M2H) collaboration and the Swope Supernova Survey as part of a search for gravitational wave counterparts on 2017 August 17 \cite{Coulter17}. The discovery image revealed a transient source with i-band magnitude of $17.476 \pm 0.018$~mag approximately 10.6 arcseconds (\arcsec) from the center of the S0-type galaxy NGC~4993 \cite{Coulter17}. We synthesize all optical and near-infrared photometry of the transient source \cite{Drout17,Shappee17}, its environment \cite{Pan17}, and our theoretical interpretation of these data.

\section*{Galactic Environment in NGC~4993}

We first consider the host galaxy of GW170817, and the location of SSS17a within it.  The host galaxy morphology and mass provide information about its age, the populations of stars that it contains, and their dynamical properties.  Comparison between different classes of astrophysical transients and their host galaxies is therefore an indirect constraint on the properties of their progenitor systems.

The host galaxy of SSS17a is NGC~4993, an S0-type galaxy with prominent lanes of cosmic dust concentrated at its core and reaching out nearly to the projected position of SSS17a \cite{Pan17}.  We analyzed the surface brightness of NGC~4993 using a {\it Hubble Space Telescope} (\hst) imaging, and found that it decreases with projected radius ($R$) and a scaling constant ($k$) as $\text{exp}(-k R^{1/4})$.  This profile is a de Vaucouleurs profile and is typical for S0-type galaxies.  Assuming this profile, we find NGC~4993 has an effective radius $R_{e} = 3.3$~kpc (17\arcsec).

Fitting stellar population models to 13-band photometry of NGC~4993 \cite{supp}, we find a galactic stellar mass of $\log (M/M_\odot) = 10.49^{+0.08}_{-0.20}$, where $M$ is the stellar mass of NGC~4993 and $M_{\odot}$ is the mass of the Sun.  This stellar mass is consistent with the host galaxies masses of SGRBs \cite{Leibler10}.

We detect no point source at the position of SSS17a in the pre-trigger \hst $F606W$ filter to a $V$-band magnitude limit of $>27.2$~mag or an absolute $V$-band magnitude of $M_{V}>-5.8$~mag at $40$~Mpc, which is the distance to NGC~4993 \cite{supp}.  This limit rules out the most luminous and massive progenitor stars, but is still consistent with the majority of massive stars \cite{Smartt15} and with low-mass stars or compact binary systems, both of which are optically dim.

SSS17a is offset 10.6\arcsec\ from the center of NGC~4993, or 2.0~kpc in projection. This corresponds to an offset of $0.6R_{e}$, which is small compared to the typical offsets seen in population studies of SGRBs \cite{Fong10}. The small offset is indicative of either a progenitor velocity less than the escape velocity of the galaxy ($\approx $350~km~s$^{-1}$ at the location of the transient; \cite{Pan17}) or a system that was recently kicked from the site where it formed.  If the latter is true, then SSS17a must be close to the site where it formed.

It is thought that the progenitor systems of SGRBs are formed more frequently in globular clusters, which have high stellar densities and more dynamical interactions per star \cite{Ramirez-Ruiz15}. There are $>$100 likely globular clusters in \hst imaging of NGC~4993, the closest offset by 290~pc in projection from SSS17a \cite{Pan17}. Even if the progenitor system was ejected from the closest cluster with a relative velocity of 10~km~s$^{-1}$, this corresponds to a travel time of 28~Myr, meaning that we cannot exclude the possibility that the SSS17a progenitor system originated in a globular cluster.

If the SSS17a progenitor system had received a large kick velocity of $>350~\text{km s}^{-1}$, it would have become unbound and exited its host galaxy after a relatively short time ($<20$~Myr). Given the lack of recent star formation seen in NGC~4993, it is unlikely that the progenitor system could have been as young as $20$~Myr. In addition, such large kicks are physically unlikely given that they tend to unbind a compact binary \cite{Chruslinska17}. Thus, it is more likely that the SSS17a progenitor was bound to its host and merged over a long timescale. This hypothesis also agrees with expectations of the progenitor systems of SGRBs in S0-type galaxies, which are thought to be systematically older and occur at lower redshifts \cite{Zheng07}.

\section*{The Optical Properties of SSS17a}

For many known classes of transients there is sufficient observational evidence to connect them to specific progenitor systems.  For instance, pre-explosion images of some supernovae reveal a massive star located at the precise position of the transient \cite{Smartt15}. Indirect evidence from optical light curves and spectroscopy also associate other transients (e.g., Type Ia and Ic supernovae) with either white-dwarf or massive star progenitors \cite{Branch95,Smartt15}.

SSS17a has different optical properties from other known astrophysical transient objects \cite{Siebert17}.  Figure~1 shows the photometric light curves, in which optical emission rises in $\lesssim$1 day, then fades rapidly, with a rapid color evolution to the red \cite{Drout17}.  The spectra shown in Figure~2 exhibit similarly rapid evolution; 11~hours after the trigger the spectrum is blue and smooth, but it transitions within a few days to a redder spectrum with at least one clear spectral bump near $9000$~\AA\ \cite{Shappee17}. These optical spectra lack the numerous  absorption features typically seen in the spectra of ordinary supernovae \cite{Shappee17,Siebert17}.

Nearly every known class of astrophysical transients is inconsistent with at least one of the following properties of SSS17a: a rise time $<11$~hours, fading with $\gsim1$~mag~day$^{-1}$, a color difference in $V$ and $H$ band magnitudes ($V-H$) that transitions from $-1.2$ to $3.6$~mag over 4~days, and the nearly featureless optical spectra at all epochs \cite{Siebert17,Drout17,Shappee17}. Among previously observed events dominated by thermal continuum emission, the most similar class is rapidly evolving blue transients \cite{Drout14}, however these events have peak magnitudes too luminous, typically occur in star-forming galaxies, do not have the observed color evolution, and do not fade as quickly as SSS17a \cite{Siebert17}. Non-thermal relativistic sources such as GRB afterglows can produce rapidly fading transients, but are not expected to produce the quasi-blackbody spectra that are observed (Figure~2) \cite{Shappee17}. 

SSS17a's observational properties suggest a different progenitor system. Various types of short duration transients have been theorized to come from white dwarf or massive star explosions, but none of these model predictions resemble SSS17a. For example, models of surface detonations on a white dwarf \cite{Bildsten07} could explain the peak luminosity and spectra dominated by thermal continuum emission.  But this model cannot explain the rapid rise time and color evolution of SSS17a detailed above. In particular, SSS17a's combination of luminosity, rise time, and red color implies a small ejecta mass ($<0.1~M_{\odot}$) consisting predominately of high-opacity, radioactive material \cite{Drout17,Siebert17}. Therefore, we turn to models of other astrophysical transients to describe our observations.

\section*{Comparison with Kilonova Models}

\paragraph*{Kilonova Models and Ultraviolet to Near-Infrared Photometry of SSS17a}

The early evolution, colors, and ultraviolet (UV) through near-infrared (IR) luminosities of SSS17a are well-constrained by its light curve \cite{Drout17}.  Observations of the optical transient began roughly 11~hours after the LIGO/Virgo trigger \cite{Coulter17}, and within 1~day of discovery the spectral energy distribution of the transient was mapped from the far-UV (\swift\ $W2$-band; $1928$~\AA) through the near-IR (Magellan/FourStar $K$-band; $21480$~\AA) \cite{supp}.

Models of compact object mergers involving neutron stars predict that optical and IR emission should be produced by radioactively powered thermal emission (a kilonova) \cite{Li98}. This emission, which is produced by sub-relativistic ejecta thrown off from the neutron stars during or just after the merger, is thought to be observable in all directions.  This physical mechanism contrasts with SGRBs, which are beamed along certain lines of sight.

Numerical simulations predict that neutron-rich material will be ejected from the system and assemble into heavy elements via rapid neutron capture ($r$-process) nucleosynthesis \cite{Lattimer74}. The optical and IR appearance of kilonovae are distinguished by the unusual composition of the ejecta. Heavy $r$-process ejecta (atomic mass $A \gtrsim 140$) include a large fraction of lanthanide elements, which have complex f-shell valence electron structures, producing millions of bound-bound line transitions \cite{Kasen13}.  This leads to a high optical opacity $\kappa \approx 5-10~\text{cm}^{2}~\text{g}^{-1}$, which causes the peak of the kilonova spectrum to be in the IR and produces a red transient lasting for days \cite{Barnes13}.  Ejecta composed primarily of lighter $r$-process products (with mass number $A \lesssim 140$) can be relatively lanthanide-free and have lower opacities, leading to a bluer transient \cite{Metzger16}.

The peak luminosity, characteristic timescale, and spectral energy distribution of a kilonova is largely determined by the total ejecta mass ($M_{\rm ej}$), characteristic ejecta velocity ($v_{\rm k}$), and lanthanide fraction in the ejecta ($X_{\rm lan}$).  To test whether our data are consistent with a kilonova, and to constrain the model parameters, we analyze the UV, optical, and IR emission of SSS17a \cite{Drout17} using numerical kilonova models \cite{Kasen17}.

The UV through IR observations of SSS17a match the numerical kilonova model (Figures~1 and 2), and indicate that the ejecta contained two distinct components.  The longer-duration IR light curves require the high opacities of a lanthanide-rich component, while the short-duration UV/optical light curves require a lower-opacity, relatively lanthanide-free component.  We model the panchromatic data by summing a red kilonova component with  $M_{\rm ej}=0.035\pm0.15~\text{M}_{\odot}$, $v_{\rm k}=0.15\pm0.03~c$, $\text{log}(X_{\rm lan})=-2.0\pm0.5$, and a blue kilonova with $M_{\rm ej}=0.025~\text{M}_{\odot}$, $v_{\rm k}=0.25~c$, and a gradient in the lanthanide fraction spanning $\text{log}(X_{\rm lan})=-4$ to $-6$ \cite{supp}. The blue kilonova reproduces both the UV luminosity and decline rate of SSS17a at early times. We do not attempt to tune the model parameters or compositional stratification.

Specific features in the data provide evidence for two-components of kilonova ejecta. The distinctive shape of the bolometric light curve  -- which declines with $\gsim1$~mag~day$^{-1}$ at times $t < 3$~days, then  plateaus from $t \approx 3 - 8$ days \cite{Drout17} -- is naturally reproduced by the sum of blue and red kilonova components. Such a spectral energy distribution can be formed by the superposition of two quasi-blackbody sources of different temperatures, and was predicted to be a signature of two-component kilonovae \cite{Barnes13}.

Around $10$--$15$~days after merger, the color temperature of SSS17a asymptotes to a value near $2500$~K \cite{Drout17}.  This behavior can be naturally explained by the recombination of open f-shell lanthanides, which occurs around this temperature\cite{Kasen13}. However, at times $> 10$~day after merger the model $H$-band luminosity is fainter than the observations. This may reflect limitations in the radiation transport model calculations, which due to uncertainties in the atomic data inputs may poorly fit the observed $H$-band luminosities. This deviation is reflected in both the derived luminosity and temperature for SSS17a \cite{Drout17}. Luminosities derived $12$~days after merger assume that the temperature remains around $2000$~K as we have no multi-band constraints beyond this point, although the model appears to be a good fit to the $K$ band data.

The characteristics of the blue kilonova are well-constrained by the early-time UV and optical detections of SSS17a. In BNS systems, it is thought that the merger scenario might involve two stars that collide violently; these dynamically expel hot polar ejecta that is neutrino irradiated and can produce a lighter $r$-process (lanthanide-free) blue kilonova \cite{Oechslin07}. The early component of SSS17a could broadly match predictions of BNS kilonovae \cite{Barnes13}.

It is thought that kilonovae from black hole/neutron star (BH-NS) systems are generated in systems with relatively small mass ratios, that is, with neutron stars  disrupted by relatively low-mass ($\lesssim5~M_{\odot}$), high-spin ($\chi_{BH}>0.8$) black holes \cite{Foucart12}. In this range, it is possible for tidal forces to eject a small fraction of the neutron star material in tidal tails, and for additional disrupted material to assemble into a disk. If the mass ratio is too high, however, the neutron star would not be disrupted until it had already plunged below the black hole's event horizon.  The dynamical ejecta in these systems should produce a lanthanide-rich, red kilonova \cite{Lattimer74}.

The photometry clearly exhibits a rapidly fading ($\gtrsim1.5$~mag~day$^{-1}$), blue kilonova in the far-UV \swift\ bands where the transient was only detected within the first day of observations (Figure~1). We find that SSS17a exhibits rapid color evolution from blue to red over the course of days \cite{Drout17}, which contrasts with that of all known optical transients, which have characteristic timescales of weeks or months \cite{Siebert17}. This rapid transformation in color is consistent, however, with the kilonovae expected from BNS merger models. 

We model the blue emission of SSS17a using a abundance gradient in the ejecta such that the lanthanide mass fraction is very low ($X_{\rm lan} = 10^{-6}$) in the outermost layers, but increases inward to a higher, though still trace, constituent ($X_{\rm lan} = 10^{-4}$) \cite{supp}. A model with such a compositional gradient provides a better fit to the color evolution of SSS17a, and is consistent with the physical expectation that the fastest moving material may have experienced the highest neutrino irradiation.

\paragraph*{Spectroscopy of SSS17a: Isolating Kilonova Features}

As a check on the validity of our kilonova models, we overplot spectra of SSS17a \cite{Shappee17} with the synthetic kilonova spectra from our kilonova model in Figure~2.  The observed spectra have been dereddened to remove the effects of dust obscuration, and the Doppler-shift caused by the recessional velocity of the host galaxy has been removed. There is qualitative agreement between the synthetic spectra and the optical emission from SSS17a, with a transition from blue at early times to red at later times \cite{Shappee17}.  Given the high velocities associated with our synthetic spectra ($\geq0.1c$) and the lanthanide features across the UV, optical, and IR, we are able to reproduce the smooth continuum shape of SSS17a with approximately the observed temperature at all epochs.

The presence of distinct blue and red thermal emission is especially apparent in the day $3$--$5$ spectra. The spectra appear to be split into two smooth continua with a transition around $9000$--$10000$~\AA, indicating that the spectrum can be divided into two components with $T>3300$~K and $T<2900$~K, which around day 3 each contribute roughly the same amount of flux.  

Our kilonova model, which is the sum of distinct blue and red eject components, reproduces the general behavior of the observed color evolution, and reproduces the characteristic double-peaked spectral continuum observed at day 3.46. Quantitatively, however, the model evolution shows several deficiencies. The blue kilonova component does not fade rapidly enough, such that the model over-predicts the optical flux at days $4.51$ and $7.45$. This suggests that the lanthanide gradient of the blue component of ejecta may be even steeper than assumed here, and the inner layers of ejecta even more lanthanide rich. This would result in a more rapid suppression of the optical flux over time. Thus, the time series of spectra provide insight into the layered compositional structure of the ejecta. 

\section*{On the Origin of $r$-Process Elements}

Theoretical studies of nucleosynthesis have discerned the thermodynamic conditions required for the $r$-process \cite{Burbidge57}. The specific astrophysical site, on the other hand, has not been unambiguously identified.

Two main alternatives have been commonly discussed. The first attributed the production of $r$-process elements to the neutron-rich regions in the outer layers of a nascent neutron star in a core-collapse supernova explosion \cite{Woosley94}, while the second suggested the ejecta from a neutron star merger as a likely site \cite{Lattimer74}. These two avenues are thought to eject different quantities of $r$-process material and the expectation is that such differences should be imprinted in the $r$-process enrichment pattern of stars in the Galactic halo \cite{Shen15}.

Because both models imply that elements are synthesized in an explosive event, a strict lower limit for the dilution of such elements can be derived by calculating the mass swept up once the blast wave has lost most of its energy to radiation. The inferred minimum $r$-process mass per event as a function of explosion energy is shown in Figure~3, which was derived using the $r$-process concentrations observed in low-metallicity stars in the Milky Way \cite{supp}. Also shown are the constraints on the ejecta mass and kinetic energy we derive for SSS17a, which demonstrate that the ejecta in this event are consistent with the most stringent constraints on $r$-process production derived using low-metallicity stars. 

The $r$-process abundance patterns in these stars (in particular elements with $A>140$) closely follow that of the Solar System, suggesting that the rate of enrichment has been relatively constant over long periods of time in Galactic history \cite{Sneden08}. The large star-to-star $r$-process dispersions that are seen in Galactic halo stars probably suggest an early and nucleosynthetically unmixed Milky Way \cite{Shen15}.

These confined heavy element inhomogeneities are then expected to be washed out as more events take place and the merger products mix effectively throughout the host galaxy. This is the case for SSS17a, which will likely deposit all of its newly synthesized heavy element-rich ejecta into NGC~4993.  

The rate of SSS17a-like events in the Galaxy is quite uncertain. Based on the observed properties of SSS17a, a conservative limit of the volumetric rate of SSS17a-like events $\Re_{\rm SSS17a} \le 9 \times 10^{-6}$~Mpc$^{-3}$~yr$^{-1}$ ($\le 9 \times 10^{3}$~Gpc$^{-3}$~yr$^{-1}$) has been derived \cite{Siebert17}.  Here we assume an average rate $\Re_{\rm SSS17a} \approx 3 \times 10^{-7}$~Mpc$^{-3}$~yr$^{-1}$ over the history of the local Universe. This value is consistent with rate limits derived from the first observing run performed by the Advanced LIGO Interferometer, although it is a factor of $3$ smaller than various predictions for what was necessary to detect a BNS merger during the second observing run \cite{Abbott16:ul}.  However, assuming 1 Milky Way-like galaxy per $(4.4{\rm Mpc})^3$ \cite{Abbott16:ul}, the assumed rate implies a Milky Way rate $R_{MW} = 25$~Myr$^{-1}$, which is consistent with predictions from binary population synthesis \cite{Chruslinska17}.

Our estimates for SSS17a show that an average total mass of about $M_{\rm r-p} \approx 0.06 M_\odot$ was ejected in the event. If this is representative of the average for BNS mergers, the cumulative mass of $r$-process elements ejected from all past SSS17a-like events over the Galactic history $t_{\rm H} \approx 10^{10}$~yr is $M_{\rm r-p} \Re_{\rm SSS17a} t_{\rm H} \approx 10^{4}$~M$_{\odot}$.

This value is comparable to the present $r$-process mass inventory in stars in the Milky Way. We determine that the total mass of $r$-process elements per Milky Way-like galaxy is $X_{\rm r-p}M_{\rm G} \approx 10^{4}$~M$_{\odot}$, where $M_{\rm G} \approx 10^{11}$~M$_{\odot}$ is the total Galactic mass in stars and gas \cite{Kafle14} and $X_{\rm r-p} \approx 10^{-7}$ is the total mass fraction of $r$-process nuclei in the Solar System \cite{Grevesse07}.  This total mass of $r$-process elements consists of 78\% light $r$-process elements ($A<140$; likely powering the blue kilonova) and and 22\% in the main component \cite{Sneden08} ($A>140$; likely powering the red kilonova). This indicates that BNS mergers such as SSS17a can produce sufficient $r$-process material to be a major source of $r$-process elements. 
 
\section*{The Progenitor System of SSS17a}

SSS17a has been observed throughout the entire electromagnetic spectrum, providing clues to the nature of the progenitor system. The radiated $\gamma$-ray flux varies on time scales of $\sim0.1$~s \cite{Goldstein17,Savchenko17}, which demonstrates that the site producing $\gamma$-ray photons must be very compact. Estimates of the total emitted energy suggest that the event must have emitted $>10^{50}$~erg \cite{Murguia-Berthier17}. Given the requirements of energy, outflow velocity and $\gamma$-ray compactness, it is unlikely that mass can be converted into outflow energy with efficiencies better than a few percent.  As a result, the central source giving rise to GW170817/GRB170817A/SSS17a must process upwards of $10^{-2}~M_{\odot}$ of material through a region which is not much larger than the size of a neutron star \cite{Lee07}.

Based on these electromagnetic constraints, several key questions remain regarding the central engine that triggered SSS17a. In particular, does the progenitor system involve a black hole and neutron star or two neutron stars? If so, can we distinguish between the two?

The UV-to-IR observations of SSS17a do not strongly constrain the primary source of this radiated emission, rather they are signatures of secondary reprocessing by the radioactive decay of $r$-process elements. From this we can determine the properties of the binary merger only by a chain of less certain inferences. However, one such inference involves the properties of the dynamical ejecta, namely its mass and velocity. These properties are predicted to correlate primarily with the mass ratio of the compact binary \cite{Korobkin12}.  

Comparing our results with numerical simulations \cite{supp} in Figure~S1, we estimate the mass ratio of the binary system to be $\sim${}$0.75$, which is consistent with findings from the gravitational wave data alone \cite{Abbott17}. This conclusion is based on a specific set of simulations, but it is generally consistent with a BNS system progenitor where the mass ratio of the progenitor is more likely to be close to 1. However, these constraints on the mass ratio depend sensitively on general-relativistic effects and the equation of state of nuclear matter \cite{Bauswein13}.

\section*{Conclusion}

SSS17a was a peculiar optical transient with a luminosity, rise time, and rate of decline unlike any known classes of optical transients \cite{Drout17,Shappee17,Siebert17}.  It occurred well within the effective radius of the massive, S0-type host galaxy NGC~4993.  Either the progenitor star system was bound to its host galaxy, or there was a $<20$~Myr delay time between the formation of the system and the optical transient.  Given the morphology of the host and the lack of any obvious signs of star formation, we infer that the former is more likely.

The light curves of SSS17a are fairly well matched by kilonova models. Such models involve the merger of a compact binary system with at least one neutron star, ejecting and synthesizing $r$-process elements.  From these models, we infer that SSS17a had an early blue kilonova component with $M_{\rm ej}=0.025~\text{M}_{\odot}$, $v_{\rm k}=0.25~c$ and a longer-lasting red kilonova with $M_{\rm ej}=0.035\pm0.15~\text{M}_{\odot}$ and $v_{\rm k}=0.15\pm0.03~c$. From this mass estimate, we find that the rate of $r$-process nucleosynthesis inferred from SSS17a is consistent with the production in our Galaxy as inferred from halo stars and the abundance of $r$-process elements in the Solar System.

From the properties of the dynamical ejecta as inferred from the UV, optical, and IR light curves of SSS17a, we infer a mass ratio of $\sim${}$0.75$ for the compact object binary progenitor.  We therefore conclude that SSS17a is most consistent with a binary neutron star progenitor system.

\bibliography{gw}

\begin{thebibliography}{10}

\bibitem{Abbott17}
B.~P. {Abbott}, {\it et~al.\/}, {\it Physical Review Letters\/} {\bf 118},
  221101 (2017).

\bibitem{Aasi15}
J.~{Aasi}, {\it et~al.\/}, {\it Classical and Quantum Gravity\/} {\bf 32},
  115012 (2015).

\bibitem{Acernese15}
F.~{Acernese}, {\it et~al.\/}, {\it Classical and Quantum Gravity\/} {\bf 32},
  024001 (2015).

\bibitem{Coulter17}
{Coulter et~al.}, {\it Science, this issue 10.1126/science.aap9811\/}  (2017).

\bibitem{Li98}
L.-X. {Li}, B.~{Paczy{\'n}ski}, {\it \apjl\/} {\bf 507}, L59 (1998).

\bibitem{Ramirez-Ruiz15}
E.~{Ramirez-Ruiz}, {\it et~al.\/}, {\it \apjl\/} {\bf 802}, L22 (2015).

\bibitem{Goldstein17}
{Goldstein et~al.}, {\it ApJL, doi:10.3847/2041-8213/aa8f41\/} {\bf 848}
  (2017).

\bibitem{Savchenko17}
{Savchenko et~al.}, {\it Astrophys. J. Letters, doi:10.3847/2041-8213/aa8f94\/}
  {\bf 848} (2017).

\bibitem{Paczynski86}
B.~{Paczynski}, {\it \apjl\/} {\bf 308}, L43 (1986).

\bibitem{Siebert17}
{Siebert et~al.}, {\it accepted to ApJL, 10.3847/2041-8213/aa905e\/}  (2017).

\bibitem{Drout17}
{Drout et~al.}, {\it Science, this issue 10.1126/science.aaq0049\/}  (2017).

\bibitem{Shappee17}
{Shappee et~al.}, {\it Science, this issue 10.1126/science.aaq0186\/}  (2017).

\bibitem{Pan17}
{Pan et~al.}, {\it submitted to ApJL, 10.3847/2041-8213/aa9116\/}  (2017).

\bibitem{supp}
Materials, methods are available as~supplementary materials.

\bibitem{Leibler10}
C.~N. {Leibler}, E.~{Berger}, {\it \apj\/} {\bf 725}, 1202 (2010).

\bibitem{Smartt15}
S.~J. {Smartt}, {\it \pasa\/} {\bf 32}, e016 (2015).

\bibitem{Fong10}
W.~{Fong}, E.~{Berger}, D.~B. {Fox}, {\it \apj\/} {\bf 708}, 9 (2010).

\bibitem{Chruslinska17}
M.~{Chruslinska}, K.~{Belczynski}, J.~{Klencki}, M.~{Benacquista}, {\it
  arXiv:1708.07885\/}  (2017).

\bibitem{Zheng07}
Z.~{Zheng}, E.~{Ramirez-Ruiz}, {\it \apj\/} {\bf 665}, 1220 (2007).

\bibitem{Branch95}
D.~{Branch}, M.~{Livio}, L.~R. {Yungelson}, F.~R. {Boffi}, E.~{Baron}, {\it
  \pasp\/} {\bf 107}, 1019 (1995).

\bibitem{Drout14}
M.~R. {Drout}, {\it et~al.\/}, {\it \apj\/} {\bf 794}, 23 (2014).

\bibitem{Bildsten07}
L.~{Bildsten}, K.~J. {Shen}, N.~N. {Weinberg}, G.~{Nelemans}, {\it \apjl\/}
  {\bf 662}, L95 (2007).

\bibitem{Lattimer74}
J.~M. {Lattimer}, D.~N. {Schramm}, {\it \apjl\/} {\bf 192}, L145 (1974).

\bibitem{Kasen13}
D.~{Kasen}, N.~R. {Badnell}, J.~{Barnes}, {\it \apj\/} {\bf 774}, 25 (2013).

\bibitem{Barnes13}
J.~{Barnes}, D.~{Kasen}, {\it \apj\/} {\bf 775}, 18 (2013).

\bibitem{Metzger16}
B.~D. {Metzger}, C.~{Zivancev}, {\it \mnras\/} {\bf 461}, 4435 (2016).

\bibitem{Kasen17}
{Kasen et~al.}, {\it submitted to Nature, doi:10.1038/nature24453\/}  (2017).

\bibitem{Oechslin07}
R.~{Oechslin}, H.-T. {Janka}, A.~{Marek}, {\it \aap\/} {\bf 467}, 395 (2007).

\bibitem{Foucart12}
F.~{Foucart}, {\it \prd\/} {\bf 86}, 124007 (2012).

\bibitem{Burbidge57}
E.~M. {Burbidge}, G.~R. {Burbidge}, W.~A. {Fowler}, F.~{Hoyle}, {\it Reviews of
  Modern Physics\/} {\bf 29}, 547 (1957).

\bibitem{Woosley94}
S.~E. {Woosley}, J.~R. {Wilson}, G.~J. {Mathews}, R.~D. {Hoffman}, B.~S.
  {Meyer}, {\it \apj\/} {\bf 433}, 229 (1994).

\bibitem{Shen15}
S.~{Shen}, {\it et~al.\/}, {\it \apj\/} {\bf 807}, 115 (2015).

\bibitem{Sneden08}
C.~{Sneden}, J.~J. {Cowan}, R.~{Gallino}, {\it \araa\/} {\bf 46}, 241 (2008).

\bibitem{Abbott16:ul}
B.~P. {Abbott}, {\it et~al.\/}, {\it \apjl\/} {\bf 832}, L21 (2016).

\bibitem{Kafle14}
P.~R. {Kafle}, S.~{Sharma}, G.~F. {Lewis}, J.~{Bland-Hawthorn}, {\it \apj\/}
  {\bf 794}, 59 (2014).

\bibitem{Grevesse07}
N.~{Grevesse}, M.~{Asplund}, A.~J. {Sauval}, {\it \ssr\/} {\bf 130}, 105
  (2007).

\bibitem{Murguia-Berthier17}
{Murguia-Berthier et~al.}, {\it Astrophys. J. Letters,
  doi:10.3847/2041-8213/aa91b3\/} {\bf 848} (2017).

\bibitem{Lee07}
W.~H. {Lee}, E.~{Ramirez-Ruiz}, {\it New Journal of Physics\/} {\bf 9}, 17
  (2007).

\bibitem{Korobkin12}
O.~{Korobkin}, S.~{Rosswog}, A.~{Arcones}, C.~{Winteler}, {\it \mnras\/} {\bf
  426}, 1940 (2012).

\bibitem{Bauswein13}
A.~{Bauswein}, S.~{Goriely}, H.-T. {Janka}, {\it \apj\/} {\bf 773}, 78 (2013).

\bibitem{LeBorgne02}
D.~{Le Borgne}, B.~{Rocca-Volmerange}, {\it \aap\/} {\bf 386}, 446 (2002).

\bibitem{Fioc97}
M.~{Fioc}, B.~{Rocca-Volmerange}, {\it \aap\/} {\bf 326}, 950 (1997).

\bibitem{Salpeter55}
E.~E. {Salpeter}, {\it \apj\/} {\bf 121}, 161 (1955).

\bibitem{Schlafly11}
E.~F. {Schlafly}, D.~P. {Finkbeiner}, {\it \apj\/} {\bf 737}, 103 (2011).

\bibitem{Cardelli89}
J.~A. {Cardelli}, G.~C. {Clayton}, J.~S. {Mathis}, {\it \apj\/} {\bf 345}, 245
  (1989).

\bibitem{Bianchi17}
L.~{Bianchi}, B.~{Shiao}, D.~{Thilker}, {\it \apjs\/} {\bf 230}, 24 (2017).

\bibitem{Skrutskie06}
M.~F. {Skrutskie}, {\it et~al.\/}, {\it \aj\/} {\bf 131}, 1163 (2006).

\bibitem{Wright10}
E.~L. {Wright}, {\it et~al.\/}, {\it \aj\/} {\bf 140}, 1868 (2010).

\bibitem{Salim16}
S.~{Salim}, {\it et~al.\/}, {\it \apjs\/} {\bf 227}, 2 (2016).

\bibitem{Clampin00}
M.~{Clampin}, {\it et~al.\/}, {\it UV, Optical, and IR Space Telescopes and
  Instruments\/}, J.~B. {Breckinridge}, P.~{Jakobsen}, eds. (2000), vol. 4013
  of {\it \procspie\/}, pp. 344--351.

\bibitem{Avila15}
R.~J. {Avila}, {\it et~al.\/}, {\it Astronomical Data Analysis Software an
  Systems XXIV (ADASS XXIV)\/}, A.~R. {Taylor}, E.~{Rosolowsky}, eds. (2015),
  vol. 495 of {\it Astronomical Society of the Pacific Conference Series\/}, p.
  281.

\bibitem{Dolphin16}
A.~{Dolphin}, {DOLPHOT: Stellar photometry}, Astrophysics Source Code Library,
  ascl:1608.013 (2016).

\bibitem{Kasen15}
D.~{Kasen}, R.~{Fern{\'a}ndez}, B.~D. {Metzger}, {\it \mnras\/} {\bf 450}, 1777
  (2015).

\bibitem{Jones09}
D.~H. {Jones}, {\it et~al.\/}, {\it \mnras\/} {\bf 399}, 683 (2009).

\bibitem{Freedman01}
W.~L. {Freedman}, {\it et~al.\/}, {\it \apj\/} {\bf 553}, 47 (2001).

\bibitem{Macias16}
P.~{Macias}, E.~{Ramirez-Ruiz}, {\it arXiv:1609.04826\/}  (2016).

\bibitem{Nishimura15}
N.~{Nishimura}, T.~{Takiwaki}, F.-K. {Thielemann}, {\it \apj\/} {\bf 810}, 109
  (2015).

\bibitem{Cowan04}
J.~J. {Cowan}, F.-K. {Thielemann}, {\it Physics Today\/} {\bf 57}, 47 (2004).

\bibitem{Rosswog13}
S.~{Rosswog}, {\it Philosophical Transactions of the Royal Society of London
  Series A\/} {\bf 371}, 20120272 (2013).

\bibitem{Rosswog17}
S.~{Rosswog}, {\it et~al.\/}, {\it Classical and Quantum Gravity\/} {\bf 34},
  104001 (2017).

\bibitem{Shen98a}
H.~{Shen}, H.~{Toki}, K.~{Oyamatsu}, K.~{Sumiyoshi}, {\it Nuclear Physics A\/}
  {\bf 637}, 435 (1998).

\bibitem{Shen98b}
H.~{Shen}, H.~{Toki}, K.~{Oyamatsu}, K.~{Sumiyoshi}, {\it Progress of
  Theoretical Physics\/} {\bf 100}, 1013 (1998).

\end{thebibliography}

\bibliographystyle{Science}

\section*{Acknowledgments}
We thank J.\ Mulchaey (Carnegie Observatories director),  L.\ Infante (Las Campanas Observatory director), and the entire Las Campanas staff for their dedication, professionalism, and excitement, all of which were critical in the observations used in this study.  We thank I.\ Thompson and the Carnegie Observatory Time Allocation Committee for approving the Swope Supernova Survey and scheduling our program.
We thank the University of Copenhagen, DARK Cosmology Centre, and the Niels Bohr International Academy for hosting D.A.C., R.J.F., A.M.B., E.R., and M.R.S.\ during the discovery of GW170817/SSS17a.  R.J.F., A.M.B., and E.R.\ were participating in the Kavli Summer Program in Astrophysics, ``Astrophysics with gravitational wave detections.''  This program was supported by the the Kavli Foundation, Danish National Research Foundation, the Niels Bohr International Academy, and the DARK Cosmology Centre.

The UCSC group is supported in part by NSF grant AST--1518052, the Gordon \& Betty Moore Foundation, the Heising-Simons Foundation, generous donations from many individuals through a UCSC Giving Day grant, and from fellowships from the Alfred P.\ Sloan Foundation (R.J.F.), the David and Lucile Packard Foundation (R.J.F.\ and E.R.) and the Niels Bohr Professorship from the DNRF (E.R.).
A.M.B.\ acknowledges support from a UCMEXUS-CONACYT Doctoral Fellowship.
D.K. is supported in part by a Department of Energy (DOE) Early Career award DE-SC0008067, a DOE Office of Nuclear Physics award DE-SC0017616, and a DOE SciDAC award DE-SC0018297, and by the Director, Office of Energy Research, Office of High Energy and Nuclear Physics, Divisions of Nuclear Physics, of the U.S. Department of Energy under Contract No.DE-AC02-05CH11231.
M.R.D.\ is a Hubble and Carnegie-Dunlap Fellow.  M.R.D.\ acknowledges support from the Dunlap Institute at the University of Toronto.
B.F.M.\ is an unpaid visiting scientist at the University of Chicago and an occasional consultant to the NASA/IPAC Extragalactic Database.
J.X.P.\ is an affiliate member of the Institute for Physics and Mathematics of the Universe.
B.J.S. acknowledges support from a Carnegie-Princeton Fellowship.
M.R.D. and B.J.S. were partially supported by NASA through Hubble Fellowship grant HST–-HF–-51373.001 awarded by the Space Telescope Science Institute.

This paper includes data gathered with the 6.5 meter Magellan Telescopes located at Las Campanas Observatory, Chile.
This research has made use of the NASA/IPAC Extragalactic Database (NED) which is operated by the Jet Propulsion Laboratory, California Institute of Technology, under contract with the National Aeronautics and Space Administration.
Based on observations made with the NASA/ESA Hubble Space Telescope, obtained from the Data Archive at the Space Telescope Science Institute (\url{https://archive.stsci.edu/hst/}), which is operated by the Association of Universities for Research in Astronomy, Inc., under NASA contract NAS 5--26555. These observations are associated with program GO--14840.

All photometry and spectroscopy of SSS17a, kilonova models, and code used in our analysis are available at \url{https://ziggy.ucolick.org/sss17a/}.

\begin{figure*}
	\begin{center}
		\includegraphics[width=0.85\textwidth]{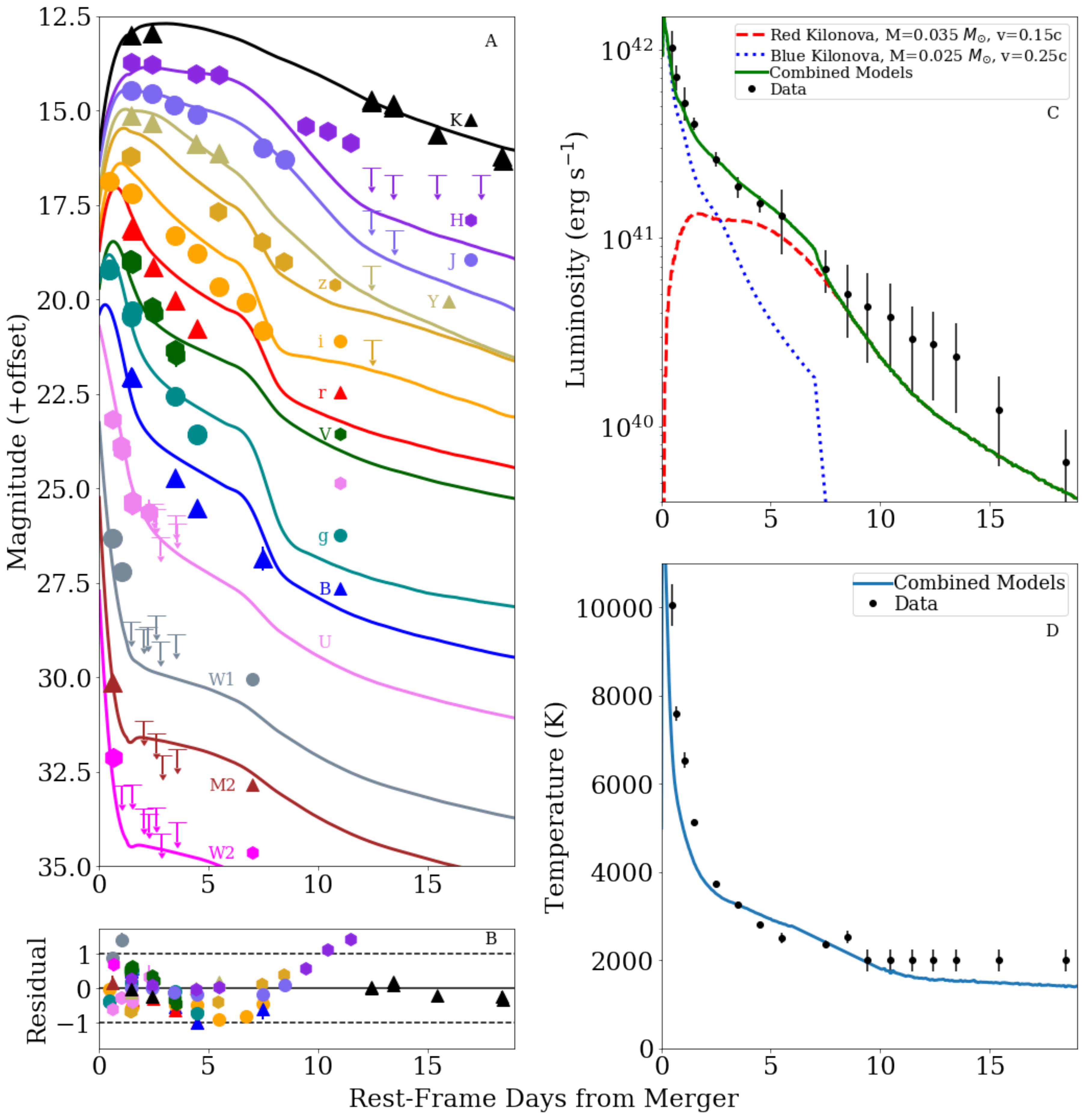}
    \end{center}
	\caption{\textbf{Photometry of SSS17a compared to fitted kilonova models}. {\it A}: UV to NIR photometry of SSS17a from $10.9$~hours after the BNS merger to +$18.5$ days \cite{Drout17}. Overplotted are our best-fitting kilonova model in each band. {\it B}: Residuals (in magnitudes) between each photometry measurement and our best-fitting model. {\it C}: The integrated luminosity of our best-fitting kilonova model compared with the total integrated luminosity of SSS17a \cite{Drout17}.  We also show the luminosity of the individual blue and red components of our kilonova model. {\it D}: The derived temperature of our kilonova model compared with the temperature derived by fitting a blackbody SED to each epoch \cite{Drout17}.}\label{fig:lightcurve}
\end{figure*}

\begin{figure}
	\begin{center}
		\includegraphics[width=0.45\textwidth]{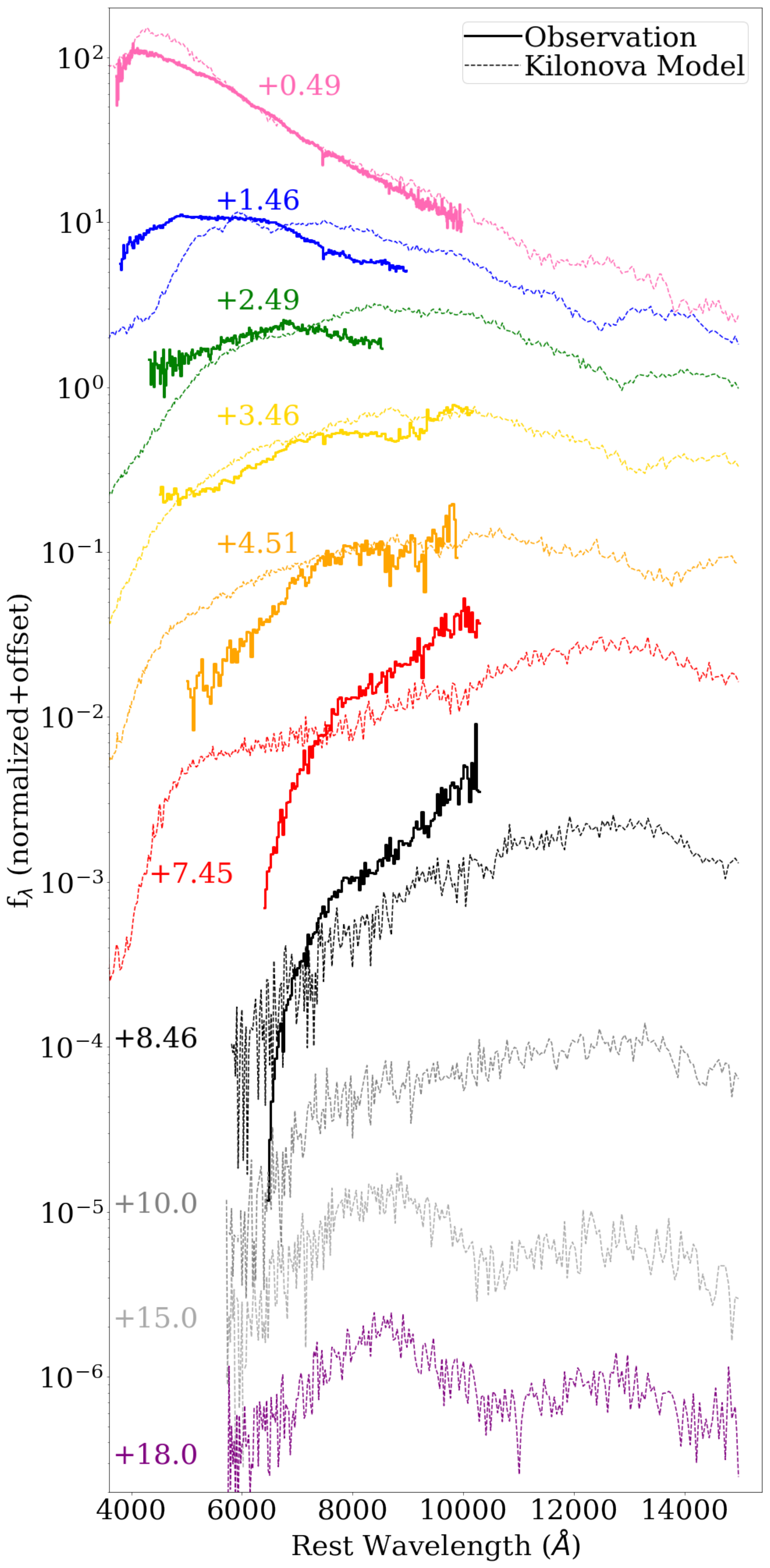}
    \end{center}
	\caption{\textbf{Spectral Series of SSS17a and Kilonova Models}. Our flux-calibrated Magellan spectra of SSS17a \cite{Shappee17}.  Each spectrum is labeled with the epoch in rest-frame days since the BNS merger.  We overplot our best-fitting kilonova model for the corresponding epoch.}\label{fig:spectrum}
\end{figure}

\begin{figure}
	\begin{center}
    	\includegraphics[width=0.7\textwidth]{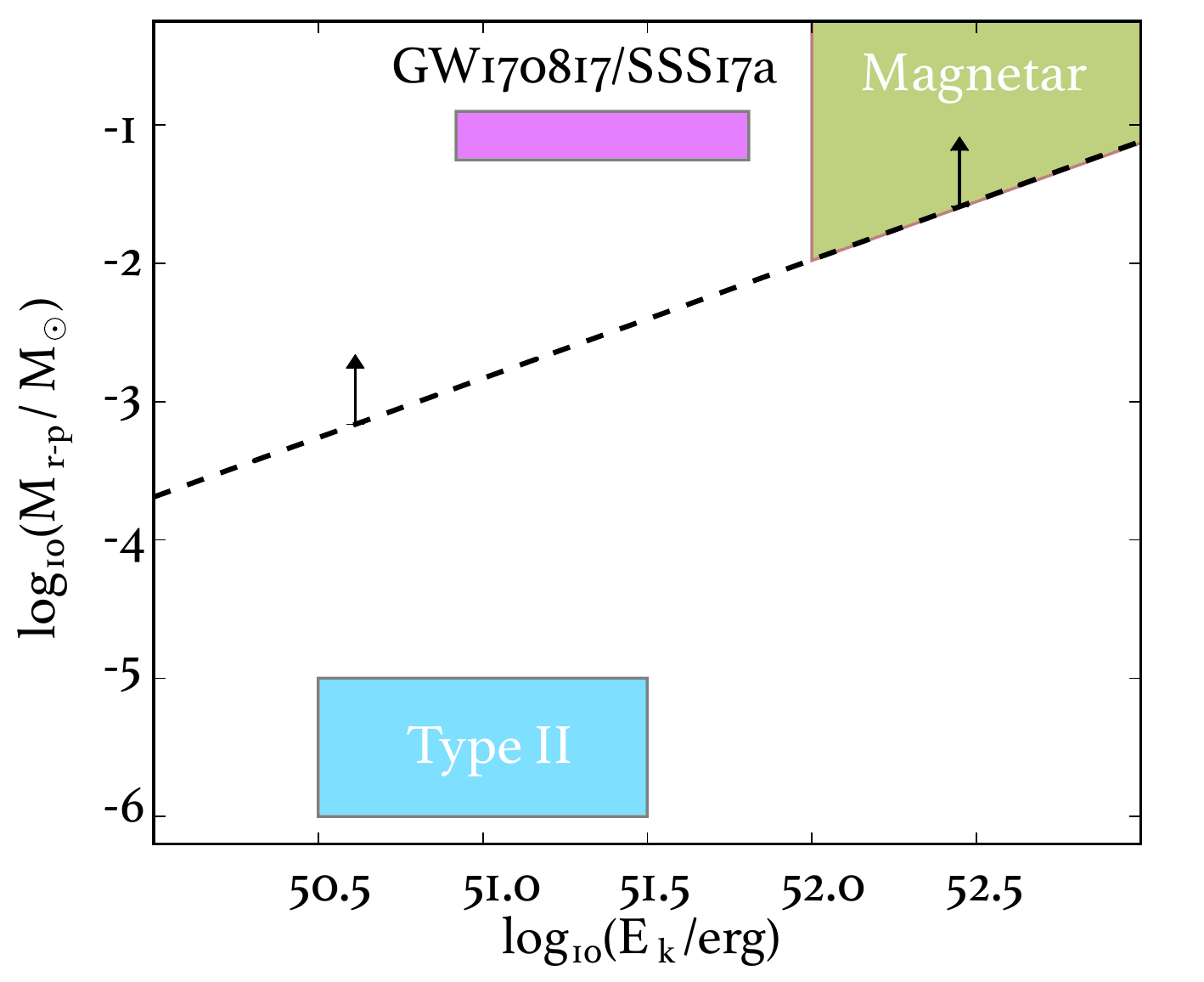}
     \end{center}
     \caption{\textbf{Mass of $r$-process Elements and Kinetic Energy in Dynamical Ejecta}. Strict lower limit on the mass ($M_{r-p}$) and kinetic energy ($E_{k}$) of $r$-process  ejecta derived from the $r$-process content of low metallicity stars (dashed line) \cite{supp}. This argument provides a range for any viable model to be tested, for example, from magnetars and Type II supernovae \cite{supp} and illustrates that SSS17a-like mergers satisfy the mass and energy ejecta constraints required for viable scenarios.}
\end{figure}

\clearpage


\clearpage
\setcounter{page}{1}
\setcounter{figure}{0}    
\renewcommand{\thefigure}{S\arabic{figure}}
\renewcommand{\thetable}{S\arabic{table}}

\begin{center}
\title{{\LARGE Supplementary Materials for}\\[0.5cm]
{\bf\large{Electromagnetic Evidence that SSS17a is the Result of a Binary Neutron Star Merger}}}

\author
{C.~D.~Kilpatrick$^{\ucsc}$,
R.~J.~Foley$^{\ucsc}$,
D.~Kasen$^{\lbl,\berk}$,
A.~Murguia-Berthier$^{\ucsc}$,\\
E.~Ramirez-Ruiz$^{\ucsc, \dark}$,
D.~A.~Coulter$^{\ucsc}$,
M.~R.~Drout$^{\car}$,
A.~L.~Piro$^{\car}$,\\
B.~J.~Shappee$^{\car,\haw}$,
K.~Boutsia$^{\lco}$,
C.~Contreras$^{\lco}$,
F.~Di~Mille$^{\lco}$,
B.~F.~Madore$^{\car}$,\\
N.~Morrell$^{\lco}$,
Y.-C.~Pan$^{\ucsc}$,
J.~X.~Prochaska$^{\ucsc}$,
A.~Rest$^{\stsci, \jhu}$,
C.~Rojas-Bravo$^{\ucsc}$,\\
M.~R.~Siebert$^{\ucsc}$,
J.~D.~Simon$^{\car}$,
N.~Ulloa$^{\nat}$
\\}

\normalsize{Correspondence to: cdkilpat@ucsc.edu}

\end{center}


\clearpage

\section{Analysis of NGC~4993}

\subsection{Photometry of NGC~4993 and Modeling}
The photometric redshift code \textsc{z-peg} \cite{LeBorgne02}, which is based on the spectral synthesis code P\'{E}GASE.2 \cite{Fioc97}, was used to estimate the stellar mass of NGC~4993 \cite{Pan17}. \textsc{z-peg} uses the P\'{E}GASE.2 stellar population synthesis code to fit the observed galaxy colors with galaxy SED templates corresponding to 9 galaxy types (SB, Im, Sd, Sc, Sbc, Sb, Sa, S0 and E). We assume a Salpeter initial-mass function (IMF) \cite{Salpeter55}. The photometry is corrected for foreground galactic extinction using a color excess $E(B-V)=0.109$\,mag \cite{Schlafly11}, and a reddening constant $R_{V}=3.1$ is assumed, which is consistent with Magellan/MIKE spectra of SSS17a \cite{Shappee17} and a standard reddening law \cite{Cardelli89}.

Using Pan-STARRS1 images of the host galaxy, we fitted an elliptical isophote to the galaxy profile using the {\tt IRAF/isophote} package to measure $griz$ AB magnitudes of $12.45\pm0.02$, $12.14\pm0.02$, $11.78\pm0.02$, and $12.62\pm0.02$\,mag, respectively. We also obtain far-ultraviolet (UV) and near-UV photometry from the Galaxy Evolution Explorer (GALEX) \cite{Bianchi17}, $JHK$ near-infrared (IR) photometry from the Two Micron All-Sky Survey (2MASS) \cite{Skrutskie06} and ch1 to ch4 IR photometry from the Wide-field Infrared Survey Explorer (WISE) \cite{Wright10,Salim16}. We measured a host stellar mass of $\text{log}(M/M_{\odot})=10.49^{+0.08}_{-0.20}$.

\subsection{\hst\,\ Imaging}

We reduced the individual frames of \hst\ Advanced Camera for Surveys \cite{Clampin00} imaging of NGC~4993 using the {\tt DrizzlePac} pipeline \cite{Avila15}.  We took the flattened, calibrated frames and corrected them for geometric distortion, sky background, and cosmic-rays before final image combination with {\tt AstroDrizzle}.  We modified the combined image World Coordinate System using {\tt TweakReg} in order to analyze the position of SSS17a \cite{Coulter17} within its host galaxy.

We performed photometry on the \hst/ACS image using {\tt dolphot} \cite{Dolphin16}.  Using the empirical image point spread function derived from {\tt dolphot}, we performed artificial star injection by inserting artificial stars of a single magnitude into the image 1000 times.  We varied the the position of each star within the instrumental PSF by drawing x,y offsets from an normal distribution centered around (x,y)=(0,0) and with a Gaussian $\sigma$ equal to the size of the instrumental PSF/$\sqrt{\text{theoretical signal-to-noise}}$.  We recovered each star and repeated the process with decreasing magnitude until we recovered $>99\%$ of all injected stars at $\geq 5~\sigma$ significance.  Using this process, we found that we could recover $>99\%$ of all stars at $\geq 5~\sigma$ significance for a $F606W$ filter magnitude $<27.2$~mag. 

\section{Optical Photometry of SSS17a}

For our UV, optical, and NIR analysis, we used the same optical photometry set presented in \cite{Drout17}, along with the Swope photometry presented in \cite{Coulter17}.  These data came from a variety of sources, including the Swope, Magellan, and du Pont telescopes as Las Campanas Observatory, \swift, the European Southern Observatory New Technology Telescope, and the Keck-I telescope.  For a full description of our data acquisition and reduction procedure, see \cite{Drout17,Coulter17}.

\section{Optical Spectroscopy of SSS17a}

We used the same set of optical spectroscopy presented in \cite{Shappee17}.  These data included Magellan/Low Dispersion Survey Spectrograph 3, Magellan Echellette Spectrograph, Magellan Inamori Kyocera Echelle, and Inamori Magellan Areal Camera and Spectrograph observations obtained between 2017 August 18 and August 25.  For a full description of our data acquisition and reduction, see \cite{Shappee17}. 

\section{Kilonova Models}

We used the synthetic kilonova models of \cite{Kasen17} following the procedures described in \cite{Kasen13,Kasen15}. These models synthesize the UV to IR spectrum of a kilonova every $0.1$~days after merger. 

Details of the model construction are given in \cite{Kasen17} and reviewed briefly here. The calculations represent a full solution of the time-dependent radiative transfer equation, which self-consistently solves for the thermal and excitation state of the ejecta under local thermodynamic equilibrium. The multi-wavelength opacities are calculated using atomic data for millions of bound-bound line transitions, including all lanthanides \cite{Kasen13,Kasen17}.

The underlying ejecta is assumed to be spherically symmetric, with the density described by a broken power law that falls off with the velocity ($v)$ as $v^{-d}$ in the inner layers with $d=1$, and more steeply with $d=10$ in the outer layers. The transition between the power-laws is set by the total mass and kinetic energy.

We compare our photometry to a modest range of models with varying ejecta mass ($M$), kinetic velocity ($v_{\rm k}$), and mass fraction of lanthanides ($X_{\rm lan}$).  Since the lanthanide species have many lines, the opacity is largely set by $X_{\rm lan}$. The range of model parameters is guided by the expected properties of the ejecta from compact object mergers, and span the range $v_{\rm k}=0.10$ to $0.40$, $M=0.01$ to $0.05$, and $\text{log}(X_{\rm lan})=-9$ to $-1$.

As discussed in \cite{Kasen17}, the early UV emission is sensitive to the density profile in the outermost layers of ejecta. To reproduce the observed UV evolution of SSS17a, we considered models with a very steep cut-off in the outer density profile ($\rho \sim v^{-50}$). This reduced the level of line blanketing in the outer tail of the ejecta, producing an appropriately bright model UV flux at early times.

We compared kilonova models to our optical photometry by forward modeling the synthetic kilonova spectra to predict the observed photometry in each epoch of our light curves.  Starting with the time-varying kilonova models, we combined single blue and red kilonova models.  The blue kilonova was constructed to match our UV and optical observations within 1~day of explosion. It had an abundance gradient in the ejecta where $X_{\rm lan} = 10^{-4} (1 + v/v_{\rm s})^{-n} + 10^{-6}$ and a characteristic velocity $v_{\rm s} = 0.32c$, and scaling exponent $n=12.0$. We independently fitted the red kilonova model to our entire set of photometry.

We added the flux from both of these models and redshifted the synthetic spectra in order to account for the recessional velocity of NGC~4993 \cite{Jones09}. We applied Milky Way-like extinction of $E(B-V)=0.106$, which matches the level of reddening toward NGC~4993 \cite{Schlafly11}.  Finally, we applied a distance modulus corresponding to $40$~Mpc \cite{Freedman01} and convolved our redshifted, reddened spectra with filter transmission functions \cite{Drout17} in order to obtain model flux densities in each filter.  We compared these model observations to our data for all times we have data.

In order to quantify the quality of the kilonova model fits, we calculated $\chi^{2}$ from the modeled flux densities and our photometric measurements and uncertainties.  Keeping the blue kilonova model fixed, our analysis used three free parameters: the mass of the red kilonova ejecta, its kinetic velocity, and the mass fraction of lanthanides.  Therefore, in addition to reporting the parameters of the best-fitting red kilonova model, we report the approximate $1$-$\sigma$ uncertainties derived for our three model parameters from $\chi^{2}$. These are $M_{\rm ej}=0.035\pm0.015~\text{M}_{\odot}$, $v_{\rm k}=0.15\pm0.03~c$, $\text{log}(X_{\rm lan})=-2.0\pm0.5$. These inferred values are subject to several systematic model uncertainties, in particular uncertainty in the radioactive heating rate, uncertainty in the ejecta opacity due to approximate atomic input data, and viewing angle effects.

\section{Constraints on the Production of $r$-process Elements}

The elemental abundances in stars residing in the Galactic halo provide constraints on the types of progenitors that merged or collapsed early in the history of the Milky Way. Of particular interest are Galactic halo stars that have iron abundances [Fe/H] from about $-2$ to $-3.5$, which are found to contain pristine $r$-process products and can be used to derive stringent constraints on how much material is required to be synthesized in a single event. We derived the total mass of $r$-process elements per Milky Way-like galaxy by estimating the total $r$-process production per solar mass of baryons from \cite{Macias16}.  The total $r$-process production in magnetars and Type II supernovae are derived from \cite{Nishimura15,Cowan04}.

\section{Constraints on the Binary Mass Ratio}

We compare our results of the properties of the dynamical ejecta with merger models. Numerical simulations \cite{Korobkin12,Rosswog13,Bauswein13} can estimate the final properties of a compact binary system. Using this set of simulations we can compare them with results from our observation in order to provide a constraint on the nature of SSS17a's progenitor. In particular, there is a predicted relation between the mass ratio and properties of the dynamical ejecta. Unequal mass ratios tend to eject more material and at higher speed \cite{Rosswog13}. Therefore, even though the models are still uncertain, we can use them to provide an estimate on the mass ratio from our observations.

We compared our observational results with a set of simulations \cite{Korobkin12,Rosswog13,Rosswog17}.  These simulations use a Smoothed Particle Hydrodynamics (SPH) code, the Shen equation of state \cite{Shen98a,Shen98b}, and consistently evolve the thermodynamical properties of the system. The predictions of these models are plotted in Figure~S1, which shows our estimates for the velocity and mass of the ejecta as a shaded region. The properties of the dynamical ejecta from the simulations \cite{Korobkin12,Rosswog13,Rosswog17} are shown  as a function of the initial mass ratio. We approximate the mass ratio of SSS17a to be $\sim{}0.75$.

\begin{figure}
	\begin{center}
		\includegraphics[width=0.8\textwidth]{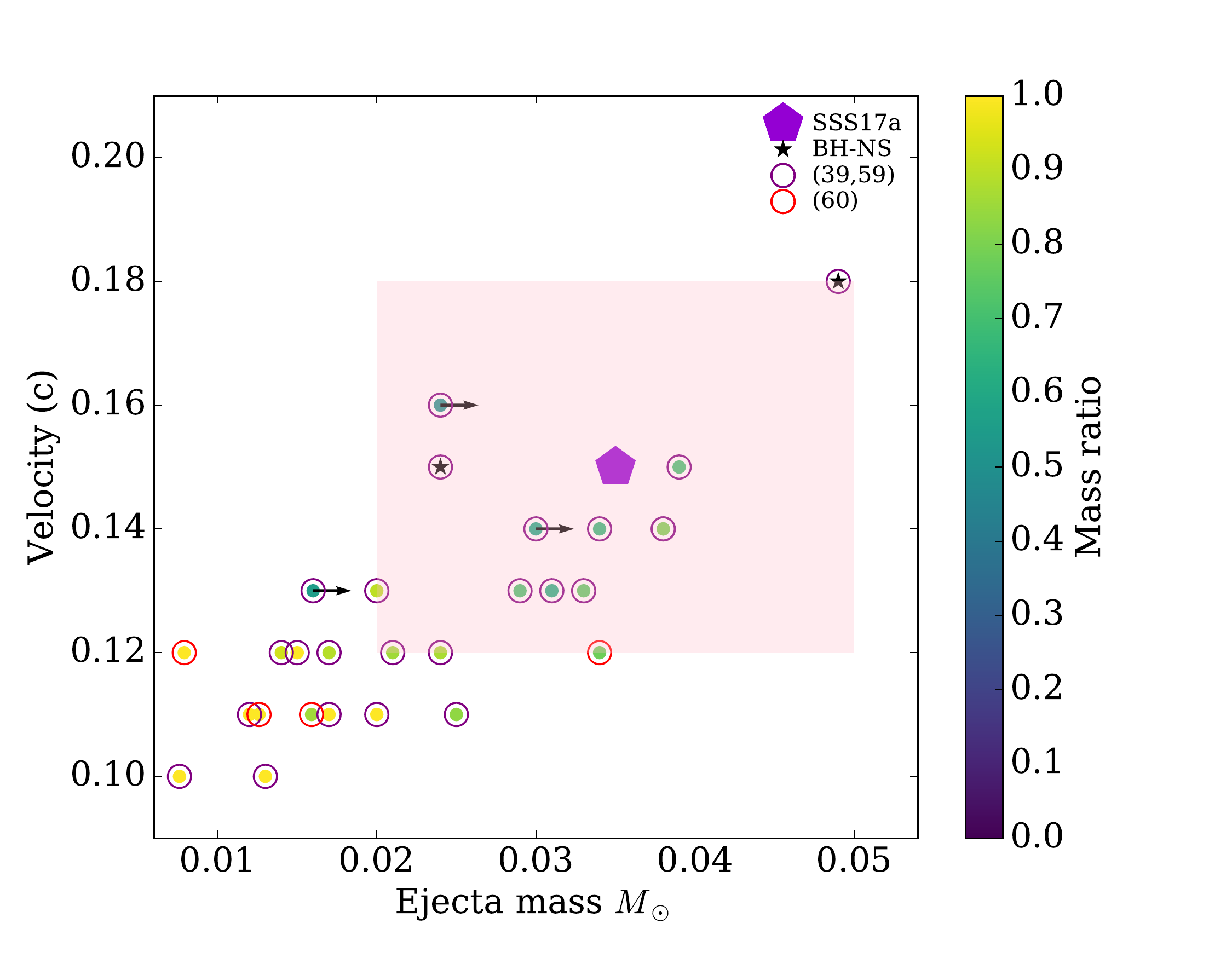}
    \end{center}
	\caption{\textbf{Mass ratio of compact binary systems as a function of the mass and velocity of the dynamical ejecta}.  These models are derived from compact binaries (BH-NS and BNS) in merger simulations \cite{Rosswog13,Korobkin12,Rosswog17}. The black stars represent BH-NS mergers, while the pentagon is the estimated value of the dynamical ejecta of SSS17a with the shaded region indicating the uncertainty of our estimates.}\label{fig:mass_ratio}
\end{figure}

\clearpage

\end{document}